\documentstyle[epsfig,aps,prd]{revtex}
\begin{document}
\draft
 \twocolumn[\hsize\textwidth\columnwidth\hsize\csname
 @twocolumnfalse\endcsname
\renewcommand{\do}{\mbox{$\partial$}}
\title{String propagation near Kaluza-Klein black holes: an analytical
and numerical study}
\author{H.~K.~Jassal\thanks{Email address : hkj@ducos.ernet.in} and
A.~Mukherjee\thanks{Email address : am@ducos.ernet.in}}
\address{Department of Physics and Astrophysics, \\
	 University of Delhi, Delhi-110 007, India.
	}
\maketitle
\begin{abstract}
This paper presents a detailed investigation of the motion of a string
near a Kaluza-Klein black hole, using the null string expansion.  
The zeroth-order string equations of motion are set up separately for
electrically and magnetically charged black hole backgrounds.  
The case of a string falling head-on into the black hole is considered
in detail. 
The equations reduce to quadratures for a magnetically charged hole,
while they are amenable to numerical solution for an electrically  
charged black hole. 
The Kaluza-Klein radius seen by the string as it approaches the black
hole decreases in the magnetic case and increases in the electric
case.  
For magnetic backgrounds, analytical solutions can be obtained in
terms of elliptical integrals. 
These reduce to elementary functions in special cases, including that
of the well-known Pollard-Gross-Perry-Sorkin monopole. 
Here the string exhibits decelerated descent into the black hole. 
The results in the authors' earlier papers are substantiated here by
presenting a detailed analysis. 
A preliminary analysis of first-order perturbations is also presented, 
and it is shown that the invariant string length receives a nonzero
contribution in the first order.   
\end{abstract}
\
\pacs{PACS: 04.50.+h, 04.70.Bw, 11.25.Db}
]
\section{Introduction}
String theories \cite{green,polchinski} possess a much richer set of
symmetries than theories of relativistic point particles.
Classically, string theory is defined in any number of dimensions, but
quantum considerations require it to be formulated in
higher dimensions - 26 for bosonic strings and 10 for superstrings.
To make a connection  with physical four-dimensional physics, the
extra dimensions are compactified.
The mechanism of this compactification is an important issue in 
string theory, and needs to be studied using a variety of
approaches.

The energy scales at which string theory is formulated are of the
order of the Planck scale. 
However, it is not unreasonable to expect that there
would be some residual effect at low energies, i.e. energies low
compared to the Planck scale, but higher than any energies 
currently accessible at particle accelerators.
One possible approach, in this spirit, could be to  study string
dynamics  in curved spacetime \cite{erice} (see also \cite{footnote1}).
The investigation is important in the context of understanding string
quantisation in curved spacetime.

The string equations of motion are complicated and various ansatze and
approximation schemes have been proposed to solve them.
One such scheme is the null string expansion
\cite{vega,vega2,sanchez}, which is suitable to describe string
propagation in strong gravitational fields.  
A possible application of this investigation could be to study the
dynamics of a string probe near a black hole where the extra
dimensions are expected to contribute nontrivially.

Solving the higher dimensional string equations of motion is
technically involved.    
An  {\it in-between} approach is provided by studying a five-dimensional
Kaluza-Klein background instead of the full $D$-dimensional spacetime 
(For a review of Kaluza-Klein theories, see \cite{overduin}). 
It would be interesting to observe how the extra
dimension appears from the point of view of a string falling into the
black hole.
An attempt in this direction was made by solving the string equations
of motion in Kaluza-Klein black hole backgrounds \cite{kkbh,brief_rep}.
It was shown that, even at the classical level, the extra compact
dimension contributes nontrivially to string propagation.
In this paper, we present a more detailed study of strings near
Kaluza-Klein black holes.
Our main tool is the null string expansion which is used to solve for
the string coordinates \cite{footnote2}.
The major results of the  zeroth order investigations have already
been reported in Refs. \cite{kkbh,brief_rep}.  
This paper provides the technical details involved in obtaining those
results; in addition, we present for the first time some analysis
of the first order equations.

In Section II,  black hole solutions to five-dimensional
Kaluza-Klein theory are reviewed.
Section III reviews string propagation in curved spacetime, with
string coordinates expanded in a perturbation series around the
worldsheet velocity of light.
Section IV is devoted to the study of string propagation in electrically
and magnetically charged  Kaluza-Klein black hole backgrounds.
Section V presents analytical results for string propagation near
extremal magnetically charged black holes.
In Section VI, the analysis of Section IV and Section V is extended up to
the first order in the  perturbation expansion and some preliminary
results presented.
The concluding Section VII contains the summary and some remarks.
Some technical details are presented in Appendix A.

\section{Kaluza-Klein Black Holes}
\renewcommand{\do}{\mbox{$\partial$}}
Stationary Kaluza-Klein solutions with spherical symmetry were studied
systematically by Chodos and Detweiler \cite{chodos} and Dobaish and
Maison \cite{dobaish}.
These black holes are characterised by the mass, the electric
charge and the scalar charge.
It was shown by Gross and Perry \cite{gross} and in independent works
by Pollard \cite{pollard} and by Sorkin \cite{sorkin} that
five-dimensional magnetic monopoles (electrically neutral) exist as
solutions to five-dimensional Kaluza-Klein theories. 
The solutions  in Ref. \cite{chodos} were  generalised by Gibbons and
Wiltshire \cite{gibbons}  to those with four parameters.  
It was shown that, in general, Kaluza-Klein black holes 
possess both electric and magnetic charge, with the solutions
mentioned above as special cases (see also \cite{miriam}).

We consider the metric background as given in  \cite{gibbons} 
\begin{eqnarray}
ds^{2}=-e^{4 k \frac{\varphi}{\sqrt{3}}} (dx_{5} &+&
2 k A_{\alpha}dx^{\alpha})^{2} \\ \nonumber
&+&e^{-2 k \frac{\varphi}{\sqrt{3}}}g_{\alpha
\beta}dx^{\alpha}dx^{\beta},
\end{eqnarray}
where $\varphi$ is the dilaton field, $k^{2}=4 \pi G$; $x_{5}$ is the extra dimension and should be
identified modulo $2\pi R_0$, where $R_0$ is the radius of the circle
about which the coordinate $x_5$ winds.

The demand that the black hole solutions be regular in
four dimensions (changing to units where $G=1$ \cite{itzhaki}) implies 
\begin{eqnarray}
e^{4 \varphi/\sqrt{3}}&=&\frac{B}{A}, \\ \nonumber
A_{\alpha}dx^{\alpha}&=&\frac{Q}{B}(r-\frac{\Sigma}{\sqrt{3}})dt+P
\cos \theta d\phi,
\end{eqnarray}
and
\begin{eqnarray}
g_{\alpha \beta}dx^{\alpha}dx^{\beta}=\frac{f^2}{\sqrt{AB}} dt^2 &-&
\frac{\sqrt{AB}}{f^2}dr^2  \\ \nonumber
&-& \sqrt{AB} \left(d \theta^2 
+ \sin^2 \theta d \phi^2\right),
\end{eqnarray}
where $A$, $B$ and $f^2$ depend on $r$ and are given by
\begin{eqnarray}
\label{eq:abf2}
A & = & (r-\frac{\Sigma}{\sqrt{3}})^{2} - \frac{2
P^{2}\Sigma}{\Sigma-\sqrt{3} M},  \\ \nonumber
B & = & (r+\frac{\Sigma}{\sqrt{3}})^{2} - \frac{2 Q^{2} \Sigma}{\Sigma
+ \sqrt{3} M},\\ \nonumber
f^{2} & = & (r-M)^{2} - (M^{2} + \Sigma^{2} - P^{2} - Q^{2}).
\end{eqnarray}
If $P=Q=0$, we regain the usual Schwarzschild black holes. 

The black hole solutions are characterised by the mass $M$ of the
black hole, the electric charge $Q$, the magnetic charge $P$ and
the scalar charge $\Sigma$.
Out of the charges, only two are independent \cite{gibbons,itzhaki}.
The constant parameters are constrained by the relation 
\begin{equation}
\frac{2}{3} \Sigma = \frac{Q^{2}}{\Sigma + \sqrt{3} M} +
\frac{P^{2}}{\Sigma - \sqrt{3} M},
\label{eq:constr}
\end{equation}
where the scalar charge is defined by  \\
\begin{center}
$k\varphi \longrightarrow \frac{\Sigma}{r} +
O\left(\frac{1}{r^2}\right)$ as $r \longrightarrow \infty.$ 
\end{center}

Eq. \ref{eq:constr} is invariant under the duality transformation
\begin{equation}
Q \longrightarrow P,~~~~P \longrightarrow Q,~~~~\Sigma \longrightarrow
-\Sigma.
\label{eq:bhduality}
\end{equation}
which relates "electric-like" and "magnetic-like" black holes.
It is worth noting that physically distinct black holes are related
by the duality.

The black hole solutions listed by Gibbons and Wiltshire include the  
much studied Pollard-Gross-Perry-Sorkin (PGPS) extremal magnetic
black hole.
Recently, it was shown that the PGPS
monopole arises as a solution of a suitable dimensionally reduced
string theory \cite{sroy}.  
This provides an additional motivation to study such black hole
backgrounds in the context of string theory.
One way is by finding out stringy corrections to the five-dimensional
black hole backgrounds \cite{itzhaki}.
A complementary approach is to study string propagation in Kaluza-Klein
black hole backgrounds, which is the subject of this paper.

\section{Null String Expansion}
We start with the bosonic string worldsheet action \cite{green} given
by   
\begin{equation}
S = -T_{0}\int d\tau d \sigma \sqrt{-det g_{ab}},
\end{equation}
where $g_{ab}=G_{\mu \nu}(X)\do_{a}X^{\mu}\do_{b}X^{\nu}$ is
the two-dimensional worldsheet metric; $\sigma$ and $\tau$ are the
worldsheet coordinates.

The classical equations of motion are given by   
\begin{equation}
\do_{\tau}^{2}X^{\mu}-c^{2}\do_{\sigma}^{2}X^{\mu}+\Gamma_{\nu
\rho}^{\mu} \left[\do_{\tau}X^{\nu}\do_{\tau}X^{\rho} -
c^{2}\do_{\sigma}X^{\nu}\do_{\sigma}X^{\rho}\right]=0,
\end{equation}
where $\Gamma_{\nu \rho}^{\mu}$ are Christoffel symbols for the
background metric.

The constraint equations are  
\begin{equation}
\do_{\tau}X^{\mu}\do_{\sigma}X^{\nu}G_{\mu \nu} = 0
\end{equation} 
\begin{equation}
[\do_{\tau}X^{\mu}\do_{\tau}X^{\nu} + c^{2}\do_{\sigma}X^{\mu}
\do_{\sigma}X^{\nu}]G_{\mu \nu} = 0. 
\label{cons2}
\end{equation}

String propagation in curved spacetime has been investigated in a number
of papers
\cite{ansatz1,ansatz2,ansatz3,ansatz4,ansatz5,ansatz6,ansatz7,ansatz8,ansatz9}
. 
Several simplifying ansatze exist to solve the highly nonlinear
string equations of motion.
We follow the approach of de Vega and Nicolaidis \cite{vega}, which uses
the  worldsheet  velocity of light  as an expansion parameter. 
The scheme involves systematic expansion in powers of $c$.
The limit of small worldsheet velocity of light corresponds to that
of small string tension. 
If $c<<1$, the coordinate expansion is suitable to describe strings in
a strong gravitational background (see \cite{vega2,sanchez}).
Here the derivatives w.r.t. $\tau$ overwhelm the $\sigma$ derivatives.  
In the opposite case ($c>>1$), the classical equations of motion give
us a stationary picture as the  $\sigma$ derivatives dominate. 
The expansion for $c=1$ corresponds to the centre of mass expansion of
the string.

We restrict ourselves to the case where $c$ is small, our interest
being to probe the dynamical behaviour of the extra dimensions.
Using this expansion scheme, the string coordinates are expressed as 
\begin{equation}
X^{\mu}(\sigma, \tau) = X_{0}^{\mu}(\sigma, \tau) + c^{2} X_{1}^{\mu}(\sigma,
\tau) + c^{4} X_{2}^{\mu}(\sigma, \tau) + 
\end{equation}
The zeroth order $X_0^{\mu}(\sigma,\tau)$ satisfies the following
set of equations:
\begin{eqnarray}
\label{eq:zero}
\ddot X_{0}^{\mu} + \Gamma_{\nu \rho}^{\mu} \dot X_{0}^{\nu} \dot
X_{0}^{\rho} &=& 0, \\ \nonumber
\dot X_{0}^{\mu} \dot X_{0}^{\nu} G_{\mu \nu} & = & 0, \\ \nonumber
\dot X_{0}^{\mu} X_{0}'^{\nu} G_{\mu \nu} & = & 0.
\end{eqnarray}

It is clear from the above equations that every point on  the string
moves along a null geodesic; thus the zeroth order  equations describe
the motion of a {\it null string}\cite{vega}. 

The first order fluctuations can be obtained by retaining terms of 
order $c^{2}$ and the equations are given by \cite{vega}
\begin{eqnarray}
\ddot{X_1}^{\rho}+2\Gamma^{\rho}_{\kappa \lambda} \dot{X_0}^{\lambda}
\dot{X_1}^{\kappa} &+& \Gamma^{\rho}_{\kappa \lambda, \alpha}
\dot{X_0}^{\kappa} \dot{X_0}^{\lambda} X_{1}^{\alpha} \\ \nonumber
&=& X_{0}''^{\rho} +
\Gamma^{\rho}_{\kappa \lambda}X_{0}'^{\kappa}X_{0}'^{\lambda}, 
\end{eqnarray}
with the constraints being
\begin{equation}
\left(2 \dot{X}_{0}^{\mu} \dot{X}_{1}^{\nu} + X_{0}'^{\mu} X_{0}'^{\nu}
\right)G_{\mu \nu} + G_{\mu \nu, \alpha} \dot{X}_{0}^{\mu} \dot{X}_{0}^{\nu}
X_{1}^{\alpha}=0, 
\label{eq:fconst}
\end{equation}
\begin{equation}
\left(\dot{X}_{1}^{\mu} X_{0}'^{\nu} + \dot{X}_{0}^{\mu} X_{1}'^{\nu}
\right) G_{\mu \nu} + G_{\mu \nu, \alpha} \dot{X}_{0}^{\mu} X_{0}'^{\nu}
X_{1}^{\alpha}=0.
\end{equation}

A physically interesting quantity is the invariant or proper string
size $l$, which is given by \cite{erice}  
\begin{equation}
d l^2 = X'^{\mu}X'^{\nu}G_{\mu \nu}(X) d\sigma^2.
\label{eq:stringsize}
\end{equation}
The differential string size has the form of an effective mass for the
geodesic motion \cite{sanchez}.
At the zeroth order, the proper string length is indeterminate. 
At the first order and higher orders, string length varies as a string
propagates in curved spacetime.

The motion of null strings in curved backgrounds has been studied in
cosmological and black hole backgrounds \cite{vega,vega2,sanchez}. 
Applying the formalism to FRW geometry, it was shown in \cite{vega}
that the string expands or contracts at the same rate as the whole
universe.  
Lousto and S\'anchez \cite{sanchez} have made an extensive study of
string propagation in conformally flat FRW spacetime and in black
hole spacetimes.  
In the next Section we discuss propagation of a null string near
Kaluza-Klein black holes.

\section{String propagation in Kaluza-Klein black hole backgrounds}

The Kaluza-Klein black hole, in general, has both magnetic and
electric charges along with a scalar charge.
Out of these three charges, two are independent as clearly shown in
Eq. (\ref{eq:constr}).
We seek to solve the equations of motion for the string coordinates in
the exterior of the black hole. 
For simplicity, we consider the magnetically and electrically charged
cases separately. 
The main results obtained in this section have been reported earlier
\cite{kkbh}; however, it contains technical details which have not been
presented before.
\subsection{Magnetically charged black hole}

The zeroth order equations of motion for the string coordinates are
obtained by substituting the above metric in Eqs. (\ref{eq:zero}).
For an electrically neutral ($Q=0$) background the equations of motion
are 
\begin{eqnarray}
\label{eoms}
\frac{\partial^{2}t}{\partial \tau^{2}} &+& 2\left(\frac{f'}{f} -
\frac{B'}{2B}\right) \frac{\partial t}{\partial \tau} \frac{\partial
r}{\partial \tau} =0, \\ \nonumber
\frac{\partial^{2}r}{\partial \tau^{2}} &+& \left[
-\frac{f^{3}}{2AB^{2}}(B'f-2f'B) \right] \left( \frac{\partial
t}{\partial \tau} \right)^{2} \\ \nonumber
&+& 
\left(\frac{A'f -  2f'A}{2Af} \right)
\left( \frac{\partial r}{\partial \tau} \right)^{2}
- 
\frac{f^{2}A'}{2A} \left( \frac{\partial \phi}{\partial \tau}
\right)^{2} \\ \nonumber
&+& 
\frac{f^2}{2A^{3}}(A'B-B'A) \left( \frac{\partial
x_{5}}{\partial \tau} \right)^{2} =0, \\ \nonumber
\frac{\partial^{2}\phi}{\partial \tau^{2}} &+& \frac{A'}{A}
\left(\frac{\partial r}{\partial \tau} \right) \left(\frac{\partial
\phi}{\partial \tau} \right) = 0, \\ \nonumber
\frac{\partial^{2} x_{5}}{\partial \tau^{2}} &+& \left(-\frac{A'}{A}
+ \frac{B'}{B} \right) \left(\frac{\partial r}{\partial
\tau}\right)\left(\frac{\partial x_{5}}{\partial \tau}\right) = 0.  
\end{eqnarray}

and the constraint equation is
\begin{eqnarray}
\frac{f^2}{B} \left(\frac{\do t}{\do \tau}\right)^2 &-& \frac{A}{f^2}
\left(\frac{\do r}{\do \tau}\right)^2 \\ \nonumber
&-& A  \left(\frac{\do \phi}{\do
\tau}\right)^2 -\frac{B}{A}  \left(\frac{\do x_5}{\do \tau}\right)^2 =0.
\label{eq:magconstr}
\end{eqnarray}

Here we have taken the string to be propagating in the equatorial
plane, i.e. $\theta=\pi/2$.
Hence the equation of motion for the coordinate $\theta$ vanishes.

The functions $A$, $B$ and $f$, for the magnetically charged black hole
case, are given by
\begin{eqnarray}
A&=&\left(r+\Sigma_{1}\right)\left(r- 3\Sigma_{1}\right), \\ \nonumber
B&=&\left(r+\Sigma_{1}\right)^2, \\ \nonumber
f^2&=&\left(r+\Sigma_{1}\right)\left(r-2M-\Sigma_{1}\right).
\end{eqnarray}
where $\Sigma_1=\Sigma/\sqrt{3}$.

The first integrals of motion are
\begin{eqnarray}
\label{firstintegs}
\frac{\do t}{\do \tau} &=&  \frac{c_{1}B}{f^{2}}, \\ \nonumber
\frac{\do \phi}{\do \tau} &=& \frac{c_{2}}{A},	 \\ \nonumber
\frac{\do x_{5}}{\do \tau} &=& c_{3}\frac{A}{B}		 \\ \nonumber
\left(\frac{\do r}{\do \tau}\right)^2 &=& \frac{B}{A}
c_{1}^2 - \frac{f^2}{A^2} c_{2}^2 - \frac{f^2}{B}c_{3}^2,
\end{eqnarray}
where $c_{1}$, $c_{2}$ and $c_{3}$ are functions of $\sigma$. 
Since, at the zeroth order, only derivatives with respect to
$\tau$ are present, we can treat $c_1$, $c_2$ and $c_3$ as constants. 
The constants $c_1$ and $c_2$ correspond respectively to the energy
$E(\sigma)$ and the angular momentum $L(\sigma)$. 
The first three equations are obtained by direct integration of the
$t$, $\phi$ and $x_5$ equations.
The constraint equation (\ref{eq:magconstr}) is then used to
obtain the equation for $\do r/\do \tau$. 

Since $A$, $B$ and $f^2$ are all functions of $r$, it is convenient to
change all the derivatives with respect to $\tau$ to those with
respect to $r$, 
\begin{eqnarray}
\frac{\do t}{\do \tau}=\frac{dt}{dr}\frac{\do r}{\do \tau},~~~~
\frac{\do x_5}{\do \tau}=\frac{dx_5}{dr}\frac{\do r}{\do \tau},~~~~
\frac{\do \phi}{\do \tau}=\frac{d\phi}{dr}\frac{\do r}{\do \tau}.
\end{eqnarray}
Here we have assumed that the trajectory can be written in the
parametric form,
\begin{equation}
t=t(r)~~~~x_5=x_5(r), ~~~~ \phi=\phi(r)
\end{equation}
 
From Eqs. (\ref{firstintegs}) we have $$\do r/\do \tau = \pm
\sqrt{\frac{B}{A} c_{1}^2 - \frac{f^2}{A^2} c_{2}^2 -
\frac{f^2}{B}c_{3}^2}$$. 
We choose the negative sign as we consider an in-falling string.

This change of variables enables us to reduce the equations to quadratures:
\begin{eqnarray}
\label{integs}
\tau &=& -\int \frac{dr}{\sqrt{\frac{B}{A}c_{1}^2 - \frac{f^2}{B}
c_{3}^2}}, \\ \nonumber 
x_{5} &=& -\int \frac{c_{3}A dr}{B\sqrt{\frac{B}{A}c_{1}^2 -
\frac{f^2}{B} c_{3}^2}}, \\ \nonumber 
t &=& -\int \frac{ c_{1} B dr}{f^2 \sqrt{\frac{B}{A}c_{1}^2 -
\frac{f^2}{B} c_{3}^2}}, 
\end{eqnarray}
up to constants of integration which depend on $\sigma$.
Here we have taken $c_{2}=0$, i.e. the string is falling in `head-on'.
The quadratures can be solved numerically to obtain $t$, $r$, and
$x_{5}$ as functions of $\tau$. 
It is clear from Eq. (\ref{eq:constr}) that, for $P^2$ to be positive,
we have
\begin{eqnarray}
\Sigma<0~~~~~{\rm or}~~~~\Sigma>\sqrt{3}M.
\end{eqnarray}

\begin{figure}
   \begin{center}
      \epsfig{file=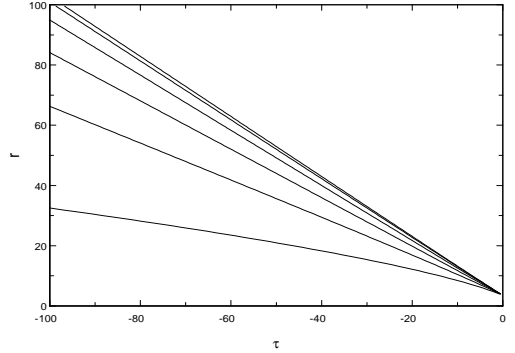, height=10.0cm, width=8.0cm}
   \end{center}
   \caption{$r$ versus $\tau$ for magnetically charged black hole.}
   \label{fig:rmag}
\end{figure}

The integrals (\ref{integs}) are evaluated numerically and inverted
to obtain the coordinates as functions of $\tau$.
We confine ourselves to the region $r>M$ and we assume $r>>\Sigma$, i.e.,
from the integrals we drop terms of $O(\Sigma^2/r^2)$.
The behaviour of $r$ as a function of $\tau$ is shown in
Fig. \ref{fig:rmag} for different values of $c_1$ and $c_3$.

\subsection{Electrically charged black hole}

For the electrically charged ($P=0$) black hole, the equations of
motion in the zeroth order take the form  
\begin{eqnarray}
\label{elec}
AB[Af^2&+&12 Q^2(r-\Sigma_{1})^2]\frac{\do^{2}t}{\do \tau^2} \\ \nonumber
&-&[12A'BQ^2(r-\Sigma_{1})^2+B'A^2f^2  \\ \nonumber 
&-&4B'AQ^2(r-\Sigma_{1})^2-2f'A^2Bf \\ \nonumber
&-&8AB Q^2 
(r-\Sigma_{1})]\frac{\do t}{\do \tau} \frac{\do r}{\do \tau} \\ \nonumber
 &+&
4QAB[B'(r-\Sigma_1)-B] \frac{\do r}{\do \tau} \frac{\do x_5}{\do
\tau} = 0,   \\ \nonumber
2A^3B^2f\frac{\do^2 r}{\do
\tau^2} &+& f^3[4A'BQ^2(r-\Sigma_1)^2-B'A^2f^2  \\ \nonumber
&+& 4B'AQ^2(r-\Sigma_1)^2  
+2f'A^2Bf \\ \nonumber
&-&8ABQ^2(r-\Sigma_1)]
\left(\frac{\do t}{\do \tau} \right)^2   \\ \nonumber
&-&
8f^3B^2Q[A'(r-\Sigma_1)-A]
\frac{\do t}{\do \tau} \frac{\do x_5}{\do \tau}  \\ \nonumber 
&+&
 A^2B^2[A'f-2f'A] \left(\frac{\do r}{\do \tau}\right)^2  \\ \nonumber
&-&A^2B^2f^3A'\left(\frac{\do \phi}{\do \tau}\right)^2 \\ \nonumber
&+&
f^3B^2[A'B-B'A] \left(\frac{\do x_5}{\do \tau}\right)^2 =0, \\ \nonumber
AB^2[Af^2&+&12Q^2(r - \Sigma_{1})^2]\frac{\do^{2}x_5}{\do
\tau^2} \\ \nonumber
&-&4QA[A'Bf^2(r-\Sigma_1) -B'Af^2(r-\Sigma_1)  \\ \nonumber
&+&
4B'Q^2(r-\Sigma_1)^3+2f'ABf(r-\Sigma_1) \\ \nonumber
&-&ABf^2 -
4BQ^2(r-\Sigma_1)^2] \frac{\do t}{\do \tau} \frac{\do r}{\do \tau} \\ \nonumber
&-&B[A'ABf^2+12A'BQ^2(r-\Sigma_1)^2 \\ \nonumber
&-&B'A^2f^2 +
4B'AQ^2(r-\Sigma_1)^2  \\ \nonumber
&-&
16ABQ^2(r-\Sigma_1)]
\frac{\do r}{\do \tau} \frac{\do x_5}{\do \tau} =0.
\end{eqnarray}
and  the constraint equation is
\begin{eqnarray}
\label{constr11}
\left\{ \frac{f^2}{B}\right.&-&\left.\frac{4Q^2}{AB}(r-\Sigma_1)^2
\right\}\left(\frac{\do t}{\do \tau} \right)^2 \\ \nonumber 
&-&\frac{8Q}{A}(r-\Sigma_1)
\frac{\do t}{\do t} \frac{\do x_5}{\do \tau} 
-\frac{A}{f^2} \left( \frac{\do r}{\do \tau} \right)^2 \\ \nonumber
&-&A \left(\frac{\do \phi}{\do \tau} \right)^2 -\frac{B}{A} \left( \frac{\do
x_5}{\do \tau} \right)^2=0,
\end{eqnarray}
where $\Sigma_1=\frac{\Sigma}{\sqrt{3}}$.

In this case, the functions $A$, $B$ and $f$ (which appear in the
metric) are
\begin{eqnarray}
A&=&(r - \Sigma_1)^2 ,\\ \nonumber
B&=& (r - \Sigma_1) (r + 3 \Sigma_1),\\ \nonumber
f^2 &=& (r - \Sigma_1) (r - 2M -\Sigma_1).
\end{eqnarray}
The structure of the equations of motion, in this case, is such that
they are  not reducible to quadratures and we have to solve the
differential equations numerically.  
Again, we consider an in-falling string in the region where $r>>\Sigma$
and $\theta=\pi/2$.

In this limit, we drop terms of $O(\Sigma_{1}^2/r^2)$ and
obtain the following equations
\begin{eqnarray}
\label{eq:eom_elec}
(r^2&-&2Mr+26\Sigma_1 M) \frac{\do^2 t}{\do
\tau^2} \\ \nonumber
&+&\frac{2(Mr+\Sigma_{1}r -12\Sigma_{1}M)}{r} \frac{\do t}{\do
\tau} \frac{\do r}{\do \tau}
+4Q \frac{\do r}{\do \tau}\frac{\do x_5}{\do
\tau} =0, \\ \nonumber
(r^2&-&2Mr+26\Sigma_1 M) \frac{\do^2 x_5}{\do
\tau^2} \\ \nonumber
&-&\frac{4Q(r+2\Sigma_{1})}{r} \frac{\do t}{\do
\tau} \frac{\do r}{\do
\tau} \\ \nonumber
&-&\frac{4\Sigma_{1}(r^2+6Mr+2\Sigma_{1}M)}{r^2} \frac{\do r}{\do
\tau}\frac{\do x_5}{\do \tau}=0,
\end{eqnarray}
with the constraint equation reducing to
\begin{eqnarray}
\label{eq:constr_elec}
\left(\frac{\do r}{\do \tau} \right)^2 &=&
\frac{1}{r^3}(r^3-4Mr^2-4M\Sigma_1r+4M^2r \\ \nonumber
&+&4M^2\Sigma_1+12M^2\Sigma_1)\left(\frac{\do
t}{\do \tau}\right)^2 \\ \nonumber
&-&\frac{8Q}{r^3}(r^2-2Mr+3\Sigma_1r-4M\Sigma_1)\frac{\do t}{\do \tau}
\frac{\do x_5}{\do \tau} \\ \nonumber
&-&
\frac{1}{r^2}(r^2+6\Sigma_1r-2Mr-10\Sigma_1M)\left(\frac{\do x_5}{\do
\tau}\right)^2.
\end{eqnarray}

The leading-order analysis of Eqs.(\ref{eq:eom_elec}) and
Eq.(\ref{eq:constr_elec}), which is required for numerical solution, is
rather involved.
Only the main results are presented here.
Details can be found in Appendix A.
To the leading order in powers of $r$, the equations can be written as
\begin{eqnarray}
\label{eq:decoupled}
r^2 \frac{d\psi_1}{dr}&-&(\Sigma_1+M)\psi_{1}^3 
+3(\Sigma_1+M)\psi_{1} \\ \nonumber
&\mp&4Q\sqrt{\psi_{1}^2-1}(\psi_{1}-1)=0, \\ \nonumber
r^2 \frac{d\psi_2}{dr}&-&6\Sigma_1 \psi_{2}^3 
+(3\Sigma_1+M)\psi_{2} \\ \nonumber
&\mp&4Q\sqrt{\psi_{2}^2+1}(\psi_{2}+1)=0,
\end{eqnarray}
where $\psi_1=\frac{dt}{dr}$, $\psi_{2}=\frac{dx_5}{dr}$
and $\psi_{1}^2 - \psi_{2}^2 =1$.
Only terms with highest power of $r$ are retained to find out the large
$r$ behaviour of the solutions.

These decoupled equations can be solved to find
\begin{eqnarray}
\label{eq:psis}
\psi_{1}\equiv \frac{dt}{dr} &=& \frac{-1}{\sqrt{{\cal A}} 
\sqrt{2M/r+c_{1}}}, \\ \nonumber
\psi_2 \equiv \frac{dx_{5}}{dr} &=& \frac{-1}{\sqrt{2M/r+c_2}},
\end{eqnarray}
where ${\cal A}$ and $c_{1}$ are constants with
$c_{1}=\frac{1-{\cal A}}{{\cal A}}$, 
and consequently, substituting in Eq. (\ref{eq:constr_elec}), we have
(see Appendix A)
\begin{equation}
\frac{\do r}{\do \tau}=I \frac{e^{-M/r}}{(2M/r+c_{1})^{1/2{\cal A}}}.
\end{equation}

Therefore, the derivatives of $t$ and $x_{5}$ w.r.t $\tau$ (in leading order)
can be written as
\begin{eqnarray}
\label{eq:dtaus}
\frac{\do t}{\do \tau} &=& \frac{-I}{\sqrt{{\cal A}} 
\sqrt{2M/r+c_{1}}}  \frac{e^{-M/r}}{(2M/r+c_{1})^{1/2{\cal A}}}
\equiv f(r), \\ 
\frac{\do x_5}{\do \tau} &=& \frac{-I}{\sqrt{{\cal A}}
\sqrt{2M/r+c_{2}}} \frac{e^{-M/r}}{(2M/r+c_{1})^{1/2{\cal A}}}
\equiv g(r).
\end{eqnarray}
The negative sign is because $r$ decreases with $\tau$ for an
in-falling string. 

The equations (\ref{eq:eom_elec}) are transformed to a convenient form
by changing variables in the following manner
\begin{eqnarray}
\frac{\do t}{\do \tau} &=& f_{1}(r) f(r),  \\ \nonumber
\frac{\do x_5}{\do \tau} &=& f_{2}(r) g(r).
\end{eqnarray}

Substituting in Eqs. (\ref{eq:eom_elec}) and defining $u=1/r$, we have
\begin{eqnarray}
\label{eq:numerical}
\frac{df_{1}}{du} &=&
\frac{-1}{(1-2Mu+26\Sigma_{1}Mu^2)}\left[-4Qf_{2}\frac{g}{f}\right. \\ \nonumber
&-&
2(M+\Sigma_{1}-12\Sigma_{1}Mu)f_{1} \\ \nonumber
&+&\left.(1-2Mu+26M\Sigma_{1}u^2)
\frac{df}{du} \frac{f_{1}}{f}\right] \\ \nonumber 
\frac{df_{2}}{du} &=&
\frac{-1}{(1-2Mu+26\Sigma_{1}Mu^2)}\left[4Q(1+2\Sigma_{1}u)f_{1}\frac{f}{g} \right. \\ \nonumber
&+&
\left.4\Sigma_{1}(1+6Mu+2\Sigma_{1}Mu^2)f_{2} \right. \\ \nonumber
&+&
\left.(1-2Mu+26M\Sigma_{1}u^2)
\frac{dg}{du} \frac{f_{2}}{g}\right]. \\ \nonumber
\end{eqnarray}

The set of Eqs. (\ref{eq:numerical}) and
Eq. (\ref{eq:constr_elec}) have been solved numerically to obtain the
coordinates as functions of $\tau$, the initial conditions for $f_{1}$
and $f_{2}$ being  
\begin{eqnarray}
f_{1}(u\rightarrow 0)\longrightarrow 1, \\ \nonumber
f_{2}(u\rightarrow 0)\longrightarrow 1.
\end{eqnarray}

\begin{figure}
   \begin{center}
      \epsfig{file=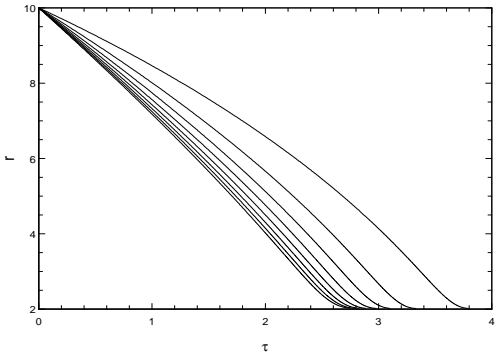, height=10.0cm, width=8.0cm}
   \end{center}
   \caption{$r$ versus $\tau$ for electrically charged black hole.}
   \label{fig:relec}
\end{figure}

Again we have a two-parameter family of solutions.
Fig. \ref{fig:relec} illustrates how $r$ varies as functions of $\tau$
for different choices of integration constants.  
Here, the constant $c_{1}$ is kept fixed at $1$ and $c_{2}$ is varied.

\subsection{Kaluza-Klein radius}

As mentioned earlier, the interest lies in seeing the effect of the extra
dimension on string propagation.
The behaviour of the extra dimension is different in the two
cases \cite{kkbh}.
However, the picture is easier to interpret if we study the
Kaluza-Klein radius, which is related to its asymptotic value $R_0$ as
\begin{equation}
R(r)  =  R_{0} \left( \frac{B}{A} \right)^{1/2}. 
\end{equation}
The radius $R(r)$ has an implicit dependence on $\tau$ through
$R(\tau)=R(r(\tau))$, and is hence a dynamical quantity.

The effect of the magnetic field is to shrink the extra dimension (as
already indicated in \cite{gibbons}), i.e., as the string approaches
the black hole, the value of the Kaluza-Klein radius which it sees
becomes smaller than its asymptotic value.  
The presence of electric charge tends to expand the extra dimension.
The opposite behaviour in the two cases was illustrated in
Ref. \cite{kkbh} and it was shown that even at the classical level,
there is a nontrivial contribution of the extra dimension on string
propagation.

\section{Analytical Results for Strings near Magnetic Black Holes}
 
\subsection{General Analytical Solutions for Magnetically Charged
Black Hole} 

The quadratures (\ref{integs}) can be solved to obtain analytical
counterparts of the solutions  presented in the previous Section for
the magnetically charged black hole. 
The integrals can be reduced to combinations of elliptical integrals,
depending on the relative values of the constants $\Sigma_1$ and $M$(see
\cite{integtab}). 
As an illustrative example we consider the case when
$c_{1}^2=c_{3}^2$. 
We can then write the integrals as 
\begin{eqnarray}
\tau&=& \frac{1}{c_{1}} \int dr \frac{\sqrt{(r - 3 \Sigma_{1})(r+\Sigma_{1})}}{\sqrt{(r
- \alpha)}}  \\ \nonumber
x_{5} &=& \int dr \frac{(r - 3 \Sigma_{1})^{3/2}} {\sqrt{(r+
\Sigma_{1}) (r - \alpha)}}\\ \nonumber
t&=& \int dr \frac{(r + \Sigma_{1})^{3/2} \sqrt{(r-3\Sigma_{1}
)}}{(r-2M-\Sigma_{1}) \sqrt{(r-\alpha)}}
\end{eqnarray}
where $\alpha= \frac{\Sigma_1 (\Sigma_1 + 3M)}{(3 \Sigma_{1} + M)}$.

As mentioned above, for $Q^2$ to be positive, the scalar charge
$\Sigma_1$ should either be negative or greater than $M$.
We first study the magnetic background with a negative scalar charge. 
We write $\Sigma_1$ as
\begin{equation}
\Sigma_1 = -e
\end{equation}
where $e$ is positive.
If $\frac{e(e-3M)}{(M-3e)}<e$, the solutions (modulo integration
constants which depend on $\sigma$) are given by 
\begin{eqnarray}
\tau &=& \frac{2}{3} \sqrt{(r+3e)(r-e) \left(r -
\frac{e(e-3M)}{(3e-M)}\right)}  \\ \nonumber
&+&\frac{16}{3}e\sqrt{e}\frac{(M+e)}{(M-3e)}
~{\rm E}\!\left(f(r),\frac{-2e}{M-3e}\right)  \\ \nonumber
&+&\frac{16}{3}e\sqrt{e}\frac{(M-e)}{(M-3e)}~{\rm F}\!\left(f(r),\frac{-2e}{M-3e}\right)
\\ \nonumber
x_{5}&=& \frac{2}{3} \sqrt{(r+3e)(r-e) \left(r -
\frac{e(e-3M)}{(3e-M)}\right)}  \\ \nonumber 
&-& \frac{8e
+\sqrt{e}(5M-11e)}{3(M-3e)}~{\rm E}\!\left(f(r),\frac{-2e}{M-3e}\right)
\\ \nonumber
&+&\frac{4e\sqrt{e}(11M-25e)}{(M-3e)}~{\rm F}\!\left(f(r),\frac{-2e}{M-3e}\right)
\\ \nonumber
t&=& \frac{2}{3} \sqrt{(r+3e)(r-e) \left(r -
\frac{e(e-3M)}{(3e-M)}\right)}  \\ \nonumber
&+&\frac{8\sqrt{e}}{3}\left[\frac{2e^2+11Me-3M^2}{M-3e}\right]~{\rm E}\!\left(f(r),\frac{-2e}{M-3e}\right)
\\ \nonumber
&-&\frac{4M}{\sqrt{e}}
{\rm \Pi} \left(-\frac{4e^2}{(M+2e)(M-3e)},f(r),\frac{-2e}{M-3e} \right) 
\end{eqnarray}
where $f(r)={\rm Arcsin}~\left[\sqrt{\frac{(r+3e)(3e-M)}{2e^2}}\right]$.
In these expressions ${\rm F}$, ${\rm E}$ and ${\rm \Pi}$ are
elliptical functions of the first, second and third kind respectively
\cite{integtab,abram}.

If $\frac{e(e-3M)}{(M-3e)}>e$, the solutions are
\begin{eqnarray}
\tau &=& \frac{2}{3} \sqrt{(r+3e)(r-e) \left(r -
\frac{e(e-3M)}{(3e-M)}\right)}  \\ \nonumber
&+&\frac{16}{3}\frac{\sqrt{2}e^2}{\sqrt{3e-M}} \frac{(M+e)}{(M-3e)} ~{\rm E}\!\left(g(r),\frac{-(M-3e)}{2e}\right)
\\ \nonumber
&+&\frac{2}{3}\frac{\sqrt{2}e^2}{\sqrt{3e-M}} \frac{(M+e)}{(M-3e)}
~{\rm F}\!\left(g(r),\frac{-(M-3e)}{2e}\right) \\ \nonumber 
x_{5}&=&  \frac{2}{3} \sqrt{(r+3e)(r-e) \left(r -
\frac{e(e-3M)}{(3e-M)}\right)}  \\ \nonumber
&-&\frac{8e\sqrt{2e}(5M-11e)}{3(M-3e)}
\frac{\sqrt{3M-5e}}{\sqrt{M-3e}}  ~{\rm E}\!\left(g(r),\frac{-(M-3e)}{2e}\right)
\\ \nonumber
&+& \frac{2e\sqrt{2e}(13M-23e)}{3(M-3e)}
~{\rm F}\!\left(g(r),\frac{-(M-3e)}{2e}\right) 
\\ \nonumber
t&=&\frac{2}{3} \sqrt{(r+3e)(r-e) \left(r -
\frac{e(e-3M)}{(3e-M)}\right)}  \\ \nonumber
&-&\frac{8\sqrt{2e}}{3\sqrt{3e-M}}(3M^2-2e^2-11Me)\\ \nonumber
&\times&~{\rm E}\!\left(g(r),\frac{-(M-3e)}{2e}\right)
\\ \nonumber
&-& \frac{4e\sqrt{2}}{3\sqrt{3e-M}}
\left[\frac{-2e(e-3M)}{(M-3e)^2}(e^2-16Me+3M^2) \right.\\ \nonumber
&+& \left.12e^2-8Me-12M^2\right]
~{\rm F}\!\left(g(r),\frac{-(M-3e)}{2e}\right) \\ \nonumber
&-&\frac{4M^2\sqrt{3e-M}}{\sqrt{2e}} {\rm \Pi} \left[ \frac{2e}{M-2e},g(r),\frac{-(M-3e)}{2e}\right],
\end{eqnarray}
where, $g(r)=\sqrt{\frac{r+3e}{4e}}$. 
The case $\Sigma_1>M$ can be treated in a very similar manner; we omit
the general expressions. 
A specific example, for the extremal black hole with $\Sigma_1=2M$ is
discussed in the next subsection.

The solutions reduce to elementary functions in the region where $r$ is
very large compared to the scalar charge.  
We take for instance the case when $c_1=c_3=1$.
Up to the first order in $\Sigma_1/r$, the solutions are 
\begin{eqnarray}
\tau &=& -\frac{2}{3 \sqrt{2(M+\Sigma_1)}} \left\{r + \frac{M
\Sigma_1}{M + 3 \Sigma_1} \right\}^{3/2} \\ \nonumber
x_5 &=& -\frac{2}{\sqrt{2(M+3\Sigma_1)}}
\left[\left(\frac{r}{3}+\alpha-\frac{9\Sigma_1}{2}\right)(r-\Sigma_1)^{1/2}\right] \\ \nonumber
&-& \frac{\left(7\Sigma_1-\alpha \right)^{3/2}}{\sqrt{2(M+\Sigma_1)}}
{\rm tan}^{-1}\left\{\frac{\sqrt{2(r-3\Sigma_1)}}{\sqrt{7\Sigma_1-\alpha}}\right\}
\\ \nonumber
t &=&
-\frac{2}{\sqrt{2(M+3\Sigma_1)}}\left[(r-\alpha)\left\{\frac{r}{3}+2M+\Sigma_1\right\}
\right] \\ \nonumber
&+&\frac{2(2M+\Sigma_1)^2}{\sqrt{2(M+3\Sigma_1)}\sqrt{\alpha-2M-\Sigma_1}}
\\ \nonumber
&\times& {\rm tan}^{-1}\left\{\frac{\sqrt{r-\alpha}}{\sqrt{\alpha-2M-\Sigma_1}}\right\}
\end{eqnarray}
where $\Sigma_1=\Sigma/\sqrt{3}$ and
$\alpha=\frac{3M\Sigma_1}{3\Sigma_1+M}$. 
These solutions are valid in the region outside the horizon but not
asymptotically far from the black hole.
The negative sign comes because we consider an in-falling string.
These solutions match with the numerical solutions presented in
the last section, for the corresponding values of $c_1$ and $c_3$
\cite{brief_rep}.

\subsection{String Equations in Extremal Black Hole backgrounds}
In the last sub-section, we discussed analytical solutions of the
equations of motion of a string propagating in a magnetically charged
black hole background.    
The solutions reduce to elementary functions in a suitable large
distance approximation, i.e., the scalar charge is very small compared
to the distance $r$.
In fact, the black hole backgrounds in this case can be thought of
as being small deviations from the Schwarzschild black holes.
The numerical results of the previous Section were also obtained in
this limit.

However, the integrals of motion (\ref{integs}) can be solved
analytically without resorting to the limit $r>>\Sigma$, if $P=2M$ and
$Q=0$, i.e, for an extremal magnetically charged black hole.
The constraint Eq. (\ref{eq:constr}) implies that $\Sigma_1=-M$ or
$\Sigma_1=2M$. 
The former case is that of the much-studied 
Pollard-Gross-Perry-Sorkin monopole \cite{pollard,gross,sorkin}.    
In that case, the metric reduces to the form  reported in
Ref.\cite{gross}. 
The solutions are
\begin{eqnarray}
\label{eq:pgps}
\tau=t &=& \frac{1}{\sqrt{c_{1}^2-c_{3}^{2}}}
(r-\beta M)^{1/2}(r+3M)^{1/2} \\ \nonumber
&+&  \frac{(3+\beta)M}{\sqrt{c_{1}^2-c_{3}^{2}}} {\rm ln}\left[(r-\beta
M)^{1/2}+(r+3M)^{1/2}\right]  \\ \nonumber
x_5 &=& \frac{1}{\sqrt{c_{1}^2-c_{3}^{2}}}
(r-\beta M)^{1/2}(r+3M)^{1/2} \\ \nonumber
&+& \frac{(11+\beta)M}{\sqrt{c_{1}^1-c_{3}^{2}}} {\rm ln}\left[(r-\beta
M)^{1/2}+(r+3M)^{1/2}\right]  \\ \nonumber
&+&\frac{16M}{\sqrt{\beta-1}\sqrt{c_{1}^2-c_{3}^{2}}}
\left[{\rm arctan}\frac{2\sqrt{r-\beta
M}}{\sqrt{\beta-1}\sqrt{r+3M}} \right]
\end{eqnarray}
where $\beta=\frac{c_{1}^2+3c_{3}^{2}}{c_{1}^2-c_{3}^{2}}$. 
We choose $c_1=1$; the condition of reality of the solutions then
forces $c_3<1$ and consequently $\beta>1$. 
Here the time coordinate $t$ is the same as the proper time
$\tau$ of the string, as in this case $f^2/B=1$.

\begin{figure}
   \begin{center}
      \epsfig{file=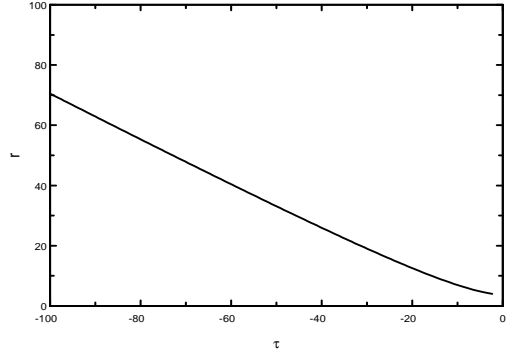, height=10.0cm, width=8.0cm}
   \end{center}
   \caption{$r$ versus $\tau$ when $\Sigma_1=-M$ for $c_1=1$ and $c_3=0.6$.}
   \label{fig:num2}
\end{figure}

Fig. \ref{fig:num2} shows a plot of $r$ versus $\tau$ for $c_3=0.6$.
A comparison of Figs. \ref{fig:rmag} and \ref{fig:num2} shows that
the approach to the horizon is different in the two cases. 
In the PGPS case, the string decelerates as it approaches the
horizon. 
This is not surprising, as the `repulsive' or `anti-gravity' effect of
extremal black holes has been commented on in the literature (see, for
example, \cite{thooft} and  \cite{gibb_BPS}). 
The effect of the gauge field is opposite to that of gravity.

In addition to the above case, there is another extremal black hole
solution (which has not been mentioned hitherto in the literature)
corresponding to $\Sigma_1=2M$. 
The integrals can be solved in terms of elliptical functions, the
solutions being  
\begin{eqnarray}
\tau &=& \frac{1}{3}\sqrt{\frac{2}{7M}} \sqrt{(r-6M)(r+2M)
\left(r-\frac{10M}{7}\right)} \\ \nonumber
&-&
\frac{32M}{21\sqrt{7}}\left[\left\{{\rm E}\!\left(g(r),\frac{3}{7}\right)
-4{\rm F}\!\left(g(r),\frac{3}{7}\right)\right\}\right] 
\\ \nonumber
x_5 &=&  \frac{1}{3}\sqrt{\frac{2}{7M}} \sqrt{(r-6M)(r+2M)
\left(r-\frac{10M}{7}\right)} \\ \nonumber
&-&
\frac{32M}{21\sqrt{7}}\left[\left\{22~{\rm E}\!\left(g(r),\frac{3}{7}\right)
-4{\rm F}\!\left(g(r),\frac{3}{7}\right)\right\} \right] \\ \nonumber 
t &=&  \frac{1}{3}\sqrt{\frac{2}{7M}}
\sqrt{(r-6M)(r+2M)\left(r-\frac{10M}{7}\right)} \\ \nonumber
&-&
\frac{2M}{21\sqrt{7}}\left[52~~{\rm F}\!\left(g(r),\frac{3}{7}\right)
\right] \\ \nonumber 
&+&\frac{M}{21\sqrt{7}}\left[59\left\{8M
~{\rm E}\!\left(g(r),\frac{3}{7}\right) 
- 6M
~{\rm F}\!\left(g(r),\frac{3}{7}\right)\right\}\right] \\ \nonumber
&+&\frac{2M}{21\sqrt{7}}\left[63~~{\rm \Pi}\!
\left(\frac{4}{7},g(r),\frac{3}{7}\right)\right],
\end{eqnarray}
where $g(r)={\rm arcsin}\left[\frac{1}{2} \sqrt{\frac{7}{6}}
\sqrt{\frac{r+2M}{M}} \right]$. 
We choose $c_1=c_3=1$.

\section{First Order Perturbations}
The effect of the background geometry on the string probe itself is
manifest at the first order and at higher orders.
For the magnetically charged black hole background, the string
equations of motion, to first order, are given by
\begin{eqnarray}
\label{eomfirst}
\ddot{B}^0&+&\Gamma^{0}_{01}
(\dot{A}^{0} \dot{B}^{1}+\dot{A}^{1}\dot{B}^{0}) 
+2\Gamma^{0}_{01,1}\dot{A}^{0}\dot{A}^{1} B^{1} \\ \nonumber
&=&A''^{0}+2\Gamma^{0}_{01} A'^{0}A'^{1} \\ \nonumber
\ddot{B}^1&+&2\Gamma^{1}_{11} \dot{A}^{1}
\dot{B}^{1}\dot{B}^{1}+2\Gamma^{1}_{55} 
\dot{A}^{5} \dot{B}^{5}
+\Gamma^{1}_{11,1} \dot{A}^{1^{2}}B^{1} \\ \nonumber
&+&\Gamma^{1}_{55,1} \dot{A}^{5^{2}} B^{1}   
=A''^{1}+\Gamma^{1}_{11}A'^{1^{2}}+\Gamma^{1}_{55}A'^{5^{2}} \\ \nonumber
\ddot{B}^5&+&2\Gamma^{5}_{51}(\dot{A}^{5}\dot{B}^{1}+\dot{A}^{1}\dot{B}^{5})
+2\Gamma^{5}_{51,1}\dot{A}^{5}\dot{A}^{1} B^{1} \\ \nonumber
&=&
A''^{5}+2\Gamma^{5}_{51} A'^{5}A'^{1}
\end{eqnarray}
and the constraints are
\begin{eqnarray}
(2\dot{A}^{0} \dot{B}^{0} &+& A''^{0^{2}}) G_{00}+ (2\dot{A}^{1}
\dot{B}^{1} + A''^{1^{2}}) G_{11} \\ \nonumber
&+&(2\dot{A}^{4} \dot{B}^{5} +
A''^{5^{2}}) G_{55}  \\ \nonumber
&+&G_{00,1}\dot{A}^{0^2} B^1 + G_{11,1} \dot{A}^{1^2} B^1 + G_{55,1}
\dot{A}^{5^2} B^1 =0. \\ \nonumber
(\dot{B}^{0} A'^{0} &+& \dot{A}^{0}B'^{0}) G_{00} + (\dot{B}^{1}
A'^{1} + \dot{A}^{1}B'^{1}) G_{11} \\ \nonumber
&+&(\dot{B}^{5} A'^{5} +
\dot{A}^{5}B'^{5}) G_{55}+G_{00,1}\dot{A}^{0}A'^{0} B^1  \\ \nonumber
&+& G_{55,1} \dot{A}^{1} A'^1 B^1 + G_{55,1}
\dot{A}^{5} A'^5 B^1 =0.
\end{eqnarray}
Here, for convenience, we have defined $A \equiv X_0$ and $B \equiv
X_1$.  
Since we confine the string to propagate in the equatorial plane and
to fall in head-on, the $B^2$ and $B^3$ equations vanish.

At the first order in $c^2$, the equations of motion are second-order
coupled partial differential equations.
The right-hand sides of Eq. (\ref{eomfirst}) involve
$\sigma$-derivatives, while the left-hand sides, which contain the
unknown functions $B^{\mu}$, involve $\tau$ derivatives.
Thus in principle the equations can be solved like ordinary differential
equations for a fixed $\sigma$.
In this Section, we consider an illustrative example without attempting to
numerically solve the equations of motion.

We take the simplest case, where $\Sigma_1 = -M$, i.e., the PGPS
monopole. 
The first order equations, in this case, are 
\begin{eqnarray}
\label{eq:fnal1}
\ddot{B}^0 = A''^{0}, \\ \nonumber
\end{eqnarray}
\begin{eqnarray}
\ddot{B}^1 &+& \frac{4M}{(r-M)(r+3M)}
\left\{\sqrt{\frac{(r-M)}{(r+3M)}c_{1}^2 - c_{3}^2}\right\}
\dot{B}^{1}  \\ \nonumber
&+& \frac{4M(r+M)}{(r-M)(r+3M)^3}c_{1}^{2} B^{1} \\ \nonumber
&-& \frac{16M^2}{(r-M)^2(r+3M)^2} c_{3}^2 B^{1}  -
\frac{4M}{(r+3M)^2}c_{3}\dot{B}^{5} 
\\\nonumber
&=& A''^{1} - \frac{2M}{(r-M)(r+3M)}A'^{1^{2}}
-\frac{2M(r-M)}{(r+3M)^3} A'^{5^{2}}, \\ \nonumber
\end{eqnarray}
\begin{eqnarray}
\ddot{B}^{5}&-&\frac{4M}{(r-M)(r+3M)}\left\{\sqrt{\frac{(r-M)}{(r+3M)}c_{1}^2
- c_{3}^2}\right\} \dot{B}^{5}  \\ \nonumber
&+&\frac{4M}{(r-M)^2}c_{3} \dot{B}^{1}  \\ \nonumber
&+&\frac{8M(r+M)}{(r-M)^{3}}c_{3}\left\{\sqrt{\frac{(r-M)}{(r+3M)}c_{1}^2
- c_{3}^2}\right\} B^{1}\\ \nonumber
&=&A''^{5}+\frac{4M}{(r-M)(r+3M)}
A'^1 A'^5, \\ \nonumber
\end{eqnarray}
with the constraint Eq. (\ref{eq:fconst}) reducing to
\begin{eqnarray}
2c_1
\dot{B}^{0}&+&\frac{2(r+3M)}{(r-M)}\left\{\sqrt{\frac{(r-M)}{(r+3M)}c_{1}^2
- c_{3}^2}\right\} \dot{B}^{1} \\ \nonumber
&-&2c_3\dot{B}^{5} 
+\frac{4M}{(r-M)(r+3M)}c_{1}^{2}B^{1} \\ \nonumber
&-&\frac{8M}{(r-M)^2} c_{3}^2 B^{1}
+A'^{0^{2}}-\frac{(r+3M)}{(r-M)}A'^{1^{2}} \\ \nonumber
&-&\frac{(r-M)}{(r+3M)}A'^{5^{2}}=0.
\end{eqnarray}
where $r=A^{1}$.

Rewriting Eqs. (\ref{eq:pgps}), including the integration
constants explicitly, we have
\begin{eqnarray}
\label{eq:pgps1}
\tau + I_1&=& \frac{1}{c_1} (A^{0}+I_0) \\ \nonumber
 &=& \frac{1}{\sqrt{c_{1}^2-c_{3}^{2}}}
(A^1-\beta M)^{1/2}(A^1+3M)^{1/2} \\ \nonumber
&+&  \frac{(3+\beta)M}{\sqrt{c_{1}^2-c_{3}^{2}}} {\rm ln}\left[(A^1-\beta
M)^{1/2}+(A^1+3M)^{1/2}\right]  \\ \nonumber
x_5 + I_{5}&=& \frac{1}{\sqrt{c_{1}^2-c_{3}^{2}}}
(A^1-\beta M)^{1/2}(A^1+3M)^{1/2} \\ \nonumber
&+& \frac{(11+\beta)M}{\sqrt{c_{1}^1-c_{3}^{2}}} {\rm ln}\left[(A^1-\beta
M)^{1/2}+(A^1+3M)^{1/2}\right]  \\ \nonumber
&+&\frac{16M}{\sqrt{\beta-1}\sqrt{c_{1}^2-c_{3}^{2}}}
\left[{\rm arctan}\frac{2\sqrt{A^1-\beta
M}}{\sqrt{\beta-1}\sqrt{A^1+3M}} \right].
\end{eqnarray}

The constant $c_1$ corresponds to the energy $E(\sigma)$. 
We choose $c_1$ and $c_3$ to be constants independent of $\sigma$.
The form of the equations depends on how the functions $I_0$, $I_{1}$
and $I_5$ depend on $\sigma$. 
We compute the derivatives of $A^0$, $A^1$ and $A^5$, viz.
\begin{eqnarray}
\frac{\do A^0}{\do \sigma} &=& -I_{0}' \\ \nonumber
\frac{\do A^1}{\do \sigma} &=& I_{1}' \sqrt{c_{1}^2 - c_{3}^2}
\frac{(A^{1}-\beta M)^{1/2}}{(A^{1}+3M)^{1/2}} \\ \nonumber
\frac{\do^2 A^1}{\do \sigma^2} &=& \sqrt{c_{1}^2 - c_{3}^2} I_{1}''
\frac{(A^{1}-\beta M)^{1/2}}{(A^{1}+3M)^{1/2}}  \\ \nonumber
&+& (c_{1}^2 - c_{3}^2)
I_{1}'^{2} \frac{(3+\beta)M}{2(A^1+3M)^2}, \\ \nonumber
\frac{\do A^5}{\do \sigma} &=& -I_{5}'+ c_3 I_{1}' 
\frac{(A^{1}+3M)} {(A^1-M)}\\ \nonumber
\frac{\do^2 A^5}{\do \sigma^2} &=& -I_{5}''+ c_3 I_{1}''
\frac{(A^{1}+3M)} {(A^1-M)}  \\ \nonumber
&-&c_3  \sqrt{c_{1}^2-c_{3}^2}
I_{1}'^{2} \frac{4M}{(A^1 -M)^2}
\frac{(A^1-\beta M)^{1/2}}{(A^1+3M)^{1/2}}.
\end{eqnarray}
The right hand sides of Eqs. (\ref{eq:fnal1}) are therefore
determined in terms of derivatives of $I_1$ and $I_5$.

For convenience, we choose $I_0'=0$, $I_1=\cos \sigma$ and $I_5 =\sin
\sigma$.  
To simplify the equations of motion further, we fix $\sigma=0$. 
Using these values of $I_1$ and $I_5$ and substituting the constraint
Eq. (\ref{eq:fconst}) in Eqs. (\ref{eq:fnal1}), we obtain
\begin{eqnarray}
\ddot{B}^1&+& \frac{c_{1}^2 4M B^1}{(r+3M)^3}=-\sqrt{c_{1}^2
-c_{3}^2} \frac{(r-\beta
M)^{1/2}}{(r+3M)^{1/2}} \\ \nonumber
&&~~~~~~~~~~~~~~~~~~-\frac{2M(r-M)}{(r+3M)^3}
\\ \nonumber
\ddot{B}^5&-&\frac{4M}{(r-M)(r+3M)}\sqrt{\frac{(r-M)}{(r+3M)} c_{1}^2 -
c_{3}^2} \dot{B}^5  \\ \nonumber
&+&\frac{4M}{(r-M)^2}c_{3}\dot{B}^1+
\frac{8M(r+M)}{(r-M)^3} c_{3} \sqrt{\frac{(r-M)}{(r+3M)} c_{1}^2 -
c_{3}^2} \\ \nonumber
&=&-c_3 \frac{(r+3M)}{(r-M)}  
\end{eqnarray}

The equations of motion are now ordinary differential equations and
can be solved to find the first order corrections to the null string
configuration. 
Although the equation for $B^1$ is now decoupled, the numerical
solution of these equations is highly nontrivial, even in the
relatively simple case of a PGPS monopole.
Moreover, in the above discussion, we have taken a particular value of
$\sigma$.
It would however be appropriate to make a more general ansatz for the
functions $I_0$, $I_1$ and $I_5$, the above example being a first
step. 

The invariant string size is defined in Eq. (\ref{eq:stringsize}) as,
\begin{equation}
d l^2 = X'^{\mu}X'^{\nu}G_{\mu \nu}(X) d\sigma^2.
\end{equation}
Using Eq. (\ref{cons2}), we have
\begin{equation}
\frac{dl^2}{d\sigma^2}=-\frac{1}{c^2}\dot{X}^{\mu} \dot{X}^{\nu} G_{\mu
\nu}. 
\end{equation}
It is clear from the above that, using the above definition, the string
length  cannot be determined in the zeroth order, i.e., for the
$c \rightarrow 0$ limit.   
The first order correction to the invariant string length is given by
\begin{equation}
\Delta \left(\frac{dl^2}{d\sigma^2} \right)=-(\dot{A}^{\mu} \dot{B}^{\nu}
+\dot{A}^{\nu} \dot{B}^{\mu} )G_{\mu \nu}  
\end{equation}
Using the constraint Eq. (\ref{eq:fconst}), the first order correction
is
\begin{equation}
\Delta \left(\frac{dl^2}{d\sigma^2} \right)=A'^{\mu}A'^{\nu} G_{\mu
\nu}(X) + G_{\mu \nu,\rho} \dot{A}^{\mu}\dot{A}^{\nu} B^{\rho}
\end{equation}
In Ref. \cite{sanchez}, the invariant string length is defined as
\begin{equation}
dl^2=-\dot{X}^{\mu} \dot{X}^{\nu} d\sigma^2,
\end{equation}
which ignores the factor $1/c^2$ and is, therefore, not appropriate to
describe null strings.

We calculate the string length for the simple case mentioned above.
The first order correction to the invariant string length for the PGPS
monopole is     
\begin{eqnarray}
\Delta \left(\frac{dl}{d\sigma}\right)^2&=& -c_{1}^2
I_{1}'^{2}\frac{(r-\beta M)}{(r-M)} - c_{3}^{2} I_{1}'^{2} \frac{(3+
\beta)M}{(r-M)}  \\ \nonumber
&+&c_{3}I_{5}'^{2}\frac{(r -\beta M)}{(r+3M)} + 2 I'_0
I'_5 \\ \nonumber
&+&\left(\frac{4M}{(r-M)(r+3M)}c_{1}^2 -  \frac{8M}{(r-M)^2}
c_{3}^2\right) B^{1}, 
\end{eqnarray}
which, for the choice of $I_1$ and $I_5$ given above and at the point
$\sigma=0$,  reduces to  
\begin{eqnarray}
\left(\frac{dl}{d\sigma}\right)^2&=& c_{3}I_{5}'^{2}\frac{(r -\beta
M)}{(r+3M)}+\left(\frac{4M}{(r-M)(r+3M)}c_{1}^2 \right. \\ \nonumber
&-& \left. \frac{8M}{(r-M)^2}
c_{3}^2\right) B^{1}. 
\end{eqnarray}
Therefore, even with the simplistic approximations described above,
there is a nontrivial contribution to the invariant string size which
can be calculated once the first order equations are solved.

\section{conclusions}
In this paper, the motion of a string near five-dimensional
Kaluza-Klein black holes has been investigated.   
For simplicity, the electrically and magnetically charged cases have been
considered separately. 
The equations of motion obtained from the string action are solved,
numerically as well as analytically, to zeroth order in the 
null string expansion.
The domain of interest is the region just outside the horizon
and in the equatorial plane. 
In the magnetically charged black hole case the equations, although
complicated, can be reduced to quadratures.
The quadratures are solved analytically in terms of different
combinations of elliptical integrals which depend on the relative
values of the charges. 
There are some solutions where the elliptical integrals reduce to
elementary functions.
Detailed analytical as well as numerical investigation is carried out
in a suitable large distance approximation, i.e. the scalar charge, in
appropriate units, being much smaller than the distance from the black
hole. 
However, for the electrically charged case, the equations cannot be
reduced to quadratures and have to be solved numerically.  
The solutions express the string coordinates in terms of the proper
time of the string. 
The extra coordinate behaves differently in the electric and magnetic
cases. 
For a clearer picture, it is useful to define a quantity called  the
Kaluza-Klein radius, which is the radius of the circle around which
the extra dimension winds. 
As the string approaches a magnetically charged black hole, the value
of the Kaluza-Klein radius which it sees becomes smaller than its
asymptotic value. 
The opposite effect is seen for an  electrically charged black hole,
where the extra dimension tends to expand.

One of the black hole solutions to five dimensional Kaluza-Klein
theory is the well known Pollard-Gross-Perry-Sorkin (PGPS)
monopole. 
It has been recently shown \cite{sroy} that this monopole arises as a
solution of a suitable  dimensionally reduced superstring theory, thus
providing an additional motivation for the present study. 
In this special case, the solutions to the string
equation of motion are obtained analytically, even though the scalar
charge is not small.  
The solution possesses an additional feature, namely, a decelerated
fall of the string into the black hole. 
This can be understood as being due to the extremal nature of the
black hole. 
The Kaluza-Klein theory possesses another extremal black hole solution
not mentioned hitherto in the literature.
The string equations of motion are solved in this background too and
the solutions written in terms of elliptical integrals.

In the first order of the null string expansion, the effects of the
background on the string  become manifest in the form of shape
changes.
The problem is technically very involved.
An illustrative calculation for the PGPS monopole case is presented to
show how one may proceed to set up the equations for numerical
solution.
Although the approximations made are simplistic, it is shown that the
invariant string size receives a nonzero contribution in the first order.
However, one should solve the second order partial differential
equations, work on which is in progress.

Our approach is complementary to finding out  stringy corrections to
the five-dimensional vacuum Einstein equations and their effect on the
black hole metrics \cite{itzhaki}. 
We limit ourselves to studying only the classical picture. 
Although, in principle, quantum effects become dominant in the strong
gravity regime, it seems plausible that the classical picture gives
an intuitive idea of the compactification mechanism.

\section*{acknowledgements}
Part of this work has been in collaboration with the late R.~P.~Saxena.
H.~K.~J. thanks Inter University Centre for Astronomy and
Astrophysics, Pune for hospitality while this manuscript was being
finalised. 
H.~K.~J. also thanks the University Grants Commission, India for
financial support.

\appendix
\section{Leading Order Analysis of Equations of Motion}

This Appendix displays some details of the leading order
analysis of string equations of motion in electrically charged
Kaluza-Klein black hole backgrounds. 
The equations of motion for an electrically charged black hole
background (Eqs. (\ref{elec})), unlike those for a magnetic black
hole, do not reduce to quadratures.
One has to therefore  resort to solving them numerically.
To find out the initial conditions for numerical solutions,
one has to analyse the equations to leading order in powers of
$1/r$.

The equations of motion, in the limit where $r>>\Sigma_1$, 
are given by Eqs. (\ref{eq:eom_elec}).
Up to the leading order, the equations take the form
\begin{eqnarray}
r^2 \frac{\do^2 t}{\do \tau^2}&+&2(M+\Sigma_1)\frac{\do t}{\do
\tau}\frac{\do r}{\do \tau} 
+4Q\frac{\do r}{\do \tau} \frac{\do x_5}{\do
\tau} =0, \\ \nonumber
r^2 \frac{\do^2 x_5}{\do \tau^2}&-&4Q\frac{\do t}{\do
\tau}\frac{\do r}{\do \tau} 
-4\Sigma_1\frac{\do r}{\do \tau}\frac{\do x_5}{\do
\tau} =0,\\ \nonumber
r^2\frac{\do^2 r}{\do \tau^2}&+&(\Sigma_1+M) \left(\frac{\do
t}{\do \tau}\right)^2
+4Q\frac{\do t}{\do \tau}\frac{\do x_5}{\do \tau}  \\ \nonumber
&+&(\Sigma_1-M) \left(\frac{\do
r}{\do \tau}\right)^2+2\Sigma_1 \left(\frac{\do
x_5}{\do \tau}\right)^2=0.
\end{eqnarray}

with the constraint
\begin{equation}
 \left(\frac{\do r}{\do \tau}\right)^2= \left(\frac{\do t}{\do
\tau}\right)^2
- \left(\frac{\do x_5}{\do \tau}\right)^2 .
\end{equation}

We make the reasonable assumption that the trajectories can be written in
the parametric form $t=t(r)$, $\theta=\theta(r)$, $\phi=\phi(r)$ and
$x_5 = x_5(r)$. 
We can then change the derivatives w.r.t. $\tau$ to  $r$ derivatives,
i.e., 
\begin{equation}
\frac{\do t}{\do \tau} =\frac{dt}{dr} \frac{\do r}{\do \tau} ,~~~~\frac{\do x_5}{\do \tau}=  \frac{dx_5}{dr}   \frac{\do r}{\do \tau}.    
\end{equation} 
We have (considering only the $t$ and $x_5$ equations for the time
being) 
\begin{eqnarray}
r^2 \frac{d^2 t}{dr^2}&-&(\Sigma_1+M) \left(\frac{dt}{dr}
\right)^3-4Q\left(\frac{dt}{dr} \right)^2\frac{dx_5}{dr} \\ \nonumber
&-&2\Sigma_1\left(\frac{dx_5}{dr} \right)^2\frac{dt}{dr}      
+(\Sigma_1+M)\frac{dt}{dr}+4Q\frac{dx_5}{dr}=0, \\ \nonumber
r^2 \frac{d^2x_5}{dr^2}&-&(\Sigma_1+M)\left(\frac{dt}{d
r}\right)^2 \frac{dx_5}{dr} -4Q\frac{dt}{dr}\left(\frac{d
x_5}{dr}\right)^2\\ \nonumber
&-&(5\Sigma_1-M)\frac{dx_5}{dr} -2\Sigma_1\left(\frac{dx_5}{d
r}\right)^3-4Q\frac{dt}{dr}=0.
\end{eqnarray}

with the constraint equation simplifying to
\begin{equation}
\left(\frac{dt}{dr}\right)^2 -\left(\frac{dx_5}{dr}\right)^2=1.
\end{equation}

Defining 
\begin{eqnarray}
\psi_1&\equiv&\frac{dt}{dr}\\ \nonumber
\psi_2&\equiv&\frac{dx_5}{dr}
\end{eqnarray}
the equations of motion are decoupled (Eqs. (\ref{eq:decoupled})) and
are given by 
\begin{eqnarray}
r^2\frac{d\psi_1}{dr}&-&
(\Sigma_1+M)\psi_{1}^3+3(\Sigma_1+M)\psi_{1}\\ \nonumber
&\mp&4Q\sqrt{\psi_{1}^2-1}(\psi_{1}-1)=0, \\ \nonumber
r^2 \frac{d\psi_2}{dr}&-&
6\Sigma_1 \psi_{2}^3+(3\Sigma_1+M)\psi_{2}\\ \nonumber
&\mp&4Q\sqrt{\psi_{2}^2+1}(\psi_{2}+1)=0,
\end{eqnarray}

These equations are first order in derivatives of $\psi_1$ and
$\psi_2$.
The equation for $\psi_1$  can be solved to
\begin{eqnarray}
&\int&\frac{d\psi_1}{(3\Sigma_1+M)\psi_{1}^3-3(\Sigma_1+M)\psi_1-4Q\sqrt{\psi_{1}^2-1}(\psi_{1}^2
-1)} \\ \nonumber
&=&-\frac{1}{r}+I_c,
\end{eqnarray}
where $I_c$ is an integration constant.

The left hand side of the equation is an integral of the type 
\begin{equation}
\int \frac{dx}{ax^3 + bx+c\sqrt{x^2-1}(x^2-1)}, 
\end{equation}
with $a=3\Sigma_1+M$, $b=-3(\Sigma_1+M)$ and $c=-4Q$. 
The electric charge $Q$ and the scalar charge $\Sigma_1$ are related
via the equation
\begin{equation}
Q^2 =\Sigma_1 (\Sigma_1 + M).
\label{eq:qq}
\end{equation}
 
With a change in variable $x={\rm sec}~\theta$ followed by the
substitution $l=\sin \theta$, and using Eq. (\ref{eq:qq}), the integral
can be written as  
\begin{equation}
\frac{1}{c} \int \frac{l dl}{l^3 - \frac{3}{2\epsilon} l^2+
\frac{1}{2\epsilon}}, 
\end{equation}
where $\epsilon =\sqrt{\frac{2\Sigma_1}{M}}$ and $\frac{b}{c}\approx
\frac{3}{4\epsilon}$ and $\frac{a+b}{c}\approx \frac{1}{2\epsilon}$.

The term $l^3$ is bounded between +1 and -1. 
We are interested in the limit of small scalar charge $\Sigma_1$,
hence $\frac{1}{\epsilon}$ is a large number. 
Therefore, the term $l^3$ in the denominator can be neglected with respect
to the terms with factors $\propto \frac{1}{\epsilon}$.
Solving the integral and resubstituting for $\psi_1$ we get the
first equation in (\ref{eq:psis}), viz.
\begin{equation}
\psi_{1}= \frac{-1}{\sqrt{{\cal A}} 
\sqrt{2M/r+c_{1}}},
\end{equation}
where the negative sign is chosen for an in-falling string.
A similar analysis for the equation for the coordinate $x_5$ gives the
expression for $\psi_2$ which is given in Eqs. (\ref{eq:psis}).

The equation of motion for $r$ (up to the leading order) is given by 
\begin{eqnarray}
\frac{\do^2 r}{\do \tau^2} &=& -\left\{\frac{\Sigma_1+M}{r^2}
\left(\frac{dt}{dr}\right)^2 \right.\\ \nonumber
&+& \left.\frac{4Q}{r^2}\frac{dt}{dr}\frac{dx_5}{dr}+\frac{\Sigma_1-M}{r^2}+\frac{2\Sigma_1}{r^2}\left(\frac{dx_5}{dr} \right)^2 \right\} \left(\frac{dr}{d\tau}\right)^2.
\end{eqnarray}

The leading order behaviour of the derivatives of $t$ and $x_5$ is
substituted in the above equation and we have
\begin{eqnarray}
\frac{\do^2 r}{\do \tau^2}&=&-\left[\frac{\Sigma_1+M}{r^2}
\frac{1}{{\cal A} (2M/r+c_1)} \right.\\ \nonumber
&+&\left.\frac{4Q}{r^2}
\frac{1}{\sqrt{{\cal A} (2M/r+c_1)(2M/r+c_2)}}\right.\\ \nonumber 
&-& \left.\frac{\Sigma_1-M}{r^2}+\frac{2\Sigma_1}{r^2} + \frac{2
\Sigma_1}{r^2} 
\frac{1}{(2M/r+c_2)} \right] \left(\frac{\do r}{\do \tau} \right)^2.
\end{eqnarray}

The first integral of motion is then given by
\begin{eqnarray}
{\rm ln}\left(\frac{\do r}{\do \tau}\right) &=&
-\left[\frac{\Sigma_1+M}{2 {\cal A}M}
{\rm ln}\left(\frac{2M}{r}+c_1\right)\right. \\ \nonumber
&+&\left. \frac{\Sigma_1}{M}
{\rm ln}\left(\frac{2M}{r}+c_2\right)+\frac{M-\Sigma_1}{r} \right] \\
\nonumber
&-&\frac{4Q}{2\sqrt{{\cal A}}M}\left[{\rm ln}~r - {\rm ln} \left(4M+c_1 r+c_2 r \right. \right.\\ \nonumber
&+& \left. \left. 2\sqrt{(2M+c_1 r)(2M+c_2 r)}\right)\right]
\end{eqnarray}

In the above equation, $\Sigma_1/M$ and
$4Q/2\sqrt{{\cal A}}M$ are of $O(\epsilon^2)$ and the terms can
therefore be neglected.
This implies,
\begin{eqnarray}
{\rm ln}\left(\frac{\do r}{\do \tau}\right) \approx -\frac{1}{2A}
{\rm ln} \left(\frac{2M}{r}+c_1\right) - \frac{M}{r},
\end{eqnarray}
which can further be solved to
\begin{equation}
\frac{\do r}{\do \tau}=\frac{I e^{-M/r}}{(2M/r+c_1)^{1/2{\cal A}}}
\end{equation}
where $I$ and $c_1$ are integration constants.
From the above values of $dt/dr$,
$dx_5/dr$ and $\do r/\do \tau$, we can obtain Eqs. (\ref{eq:dtaus}).

This equips us with the initial conditions for the numerical solutions
of Eqs. (\ref{eq:eom_elec}).
These equations can be solved to find the behaviour of $t$ and $x_5$ 
and the constraint equation can then be used to find out $r$ as a
function of $\tau$.


\begin{references}
\bibitem{footnote2}Although motivated by fundamental
string theory, these methods are equally applicable to cosmic strings
moving in black hole backgrounds as already mentioned in
\cite{vega,vega2,sanchez}.
\end{references}
\end{document}